\documentclass[graybox]{svmult}
\usepackage{aas_macros}
\usepackage{type1cm}        
\usepackage{makeidx}         
\usepackage{graphicx}        
\usepackage{multicol}        
\usepackage[bottom]{footmisc}
\usepackage{soul}
\usepackage{xspace}
\usepackage{newtxtext}       %
\usepackage{newtxmath}       
\usepackage{comment}
\usepackage{subcaption}
\usepackage{amsmath}
\usepackage[normalem]{ulem}
\def \hcm {\hbox {\ifmmode $ atom cm$^{-2}\else atom cm$^{-2}$\fi}}

\def\apjl{ApJ, }
\def\apjs{ApJS, }
\def \aap {A\&A, }

\def\ltsima{$\; \buildrel < \over \sim \;$}
\def\simlt{\lower.5ex\hbox{\ltsima}}
\def\gtsima{$\; \buildrel > \over \sim \;$}
\def\simgt{\lower.5ex\hbox{\gtsima}}

\def\psr1023{PSR J1023+0038}

\def\3fgl{3FGL J1544.6-1125}
\def\j0838{3FGL J0838.8-2829}

\newcommand{\be}{\begin{equation}}
\newcommand{\ee}{\end{equation}}

\makeindex             

\begin{document}

\title*{Millisecond Magnetars}
\author{Simone Dall'Osso 
\& Luigi Stella}
\institute{Simone Dall'Osso\at Gran Sasso Science Institute, Viale Crispi 7, 67100, L'Aquila, Italy, \email{simone.dallosso@gssi.it} \\
INFN - Laboratori Nazionali del Gran Sasso, I-67100, L’Aquila, Italy\\
\\
Luigi Stella\at INAF -- Osservatorio Astronomico di Roma, Via di Frascati 33, Monte Porzio Catone (Roma), I-00078, Italy, \email{luigi.stella@inaf.it}}

%
%
\maketitle
\abstract{Two classes of X-ray/gamma-ray sources, the Soft Gamma Repeaters and the 
Anomalous X-ray Pulsars have been identified with isolated, slowly spinning {\it magnetars}, neutron stars
whose emission draws energy from their extremely strong magnetic field ($\sim10^{15}-10^{16}$~G). Magnetars are believed to form with millisecond spin period and to represent an important fraction 
of the whole population of young neutron stars. Newborn magnetars can convert very quickly their rotational energy into electromagnetic and/or gravitational waves, by virtue of their strong magnetic fields and fast spins. This chapter provides a brief summary of astrophysical problems and scenarios in which millisecond magnetars are believed to play a 
key role: these include Gamma Ray Bursts, Supernovae, Gravitational Wave events and Fast Radio Bursts.} 

\section{Introduction} 
\label{sec:1}
The idea that neutron stars (NSs) may possess surface magnetic fields in excess of $B_{\rm crit} = 4.3 \times 10^{14}$~G, the field
for which the cyclotron energy equals the electron rest mass, predates the discovery of radio pulsars\cite{Woltjer64}. 
Substantially lower magnetic fields, typically in the $\sim 10^{11}-10^{13}$~G range, were derived in the late 60's and 70's
from the application of the rotating magnetic dipole model to the spin-down of classical radio pulsars and 
the measurement of cyclotron resonant scattering features in the X-ray spectra of accreting pulsars\cite{OstrikerG69, TrumperEA78}. 
Supra-critical B-fields did not receive much attention for many years to follow. The idea was  
resurrected and developed especially in interpreting the properties of two classes of high energy sources,
the Soft Gamma Repeaters (SGRs; \cite{KouveliotouEA87})
and the Anomalous X-ray Pulsars (AXPs; \cite{MereghettiS95}). The former~were identified after the soft, sub-second long $\gamma$-ray flares emitted~by the proto-typical~class member SGR 0526-66 \cite{Maz79}, while the latter initially comprised~the four persistent X-ray pulsars 1E2259+586, 1E 1048.1-5937, 4U0142+61 and RX J1838.4-0301, with spin periods $\sim 5-10$ s, X-ray luminosities largely in excess of their putative spindown power and no signs of companion stars from which they may accrete material \cite{MereghettiS95, VanTaamVan95}. The subsequent detection of persistent X-ray pulsations from SGRs, with similar spin period, spindown power and X-ray luminosity as AXPs (e.g. SGR 1806-20; \cite{Kou98}), and SGR-like flares from AXPs (e.g. 1E2259+586;  \cite{GaKas02}), demonstrated the close relation between the two classes.

A few tens SGRs and AXPs are presently known in our Galaxy (plus one in the 
LMC); they are isolated X-ray/soft gamma-ray sources 
displaying pulsations at relatively slow spin periods of 
$\sim 1 - 12$~s and with spin-down ages of $\sim 240 -10^5$~yr \footnote{for the lower limit of this range see \cite{EspositoEA20}.}.
Their association to supernova remnants in about one third of the cases  
and location close to the  galactic plane 
further testify to their very young age. Some 
are persistent X-ray sources 
(typical luminosities in the $10^{34} - 10^{35}$~erg/s range) and  others 
are transients, which spend most of the time in quiescence at very low 
luminosity levels 
and undergo month/year-long outbursts during which their properties closely
resemble those of persistent AXPs and SGRs.
The emission of subsecond duration X-ray bursts with peak luminosities of 
$\sim 10^{38} - 10^{41}$~erg/s during sporadic 
periods of activity is a defining characteristic of both classes. On rare 
occasions more extreme events known as intermediate and 
giant flares, lasting from several second to minutes, have been 
observed which involve much larger energy releases, up to $\sim 10^{44} - 10^{46}$~ergs.
Occasionally radio pulsations at the spin period or individual radio bursts 
have been observed in a few AXPs and SGRs. 

In many cases the measured spin-down rate implies rotational energy losses that are about two 
orders of magnitude smaller  than the $\sim 10^{34}-10^{35}$ erg/s persistent X-ray 
luminosity. Therefore the main source of power of AXPs/SGRs cannot be rotation: that is 
unlike classical radio pulsars. The highly super-Eddington luminosity of sub-second bursts and, even more so, giant flares rules out accretion as their powering mechanism. 
Reviews of the  properties of AXPs/SGRs are found  
{\it e.g.} in \cite{Mereghetti05, WoodT06, ReaE11}.

In a series of papers Robert Duncan and Chris Thompson introduced the magnetar 
model and proposed that the emission of AXPs and SGRs is generated (mostly) at the expense 
of their magnetic energy \cite{DuncanT92, ThompsonD93, ThompsonD95, ThompsonD96}. The model\footnote{for a basic tutorial see http://solomon.as.utexas.edu/magnetar.html}  
envisages that the external magnetic field 
of magnetars drives angular momentum losses via magnetic dipole radiation 
and pulsar wind as in rotation-powered pulsars, whereas the energy release 
resulting from instabilities of their inner, mainly toroidal  
magnetic field is responsible for the bulk of the emission. 
According to the magnetar model energy is fed to the magnetosphere
as magnetic field's helicity propagates~from~the interior outwards. 
Impulsive high-energy bursts originate in sudden energy injections 
resulting from {\it crust-quakes}; persistent emission, in turn, arises 
from the gradual untwisting of the magnetosphere. 
Giant flares are believed to originate 
from large-scale rearrangements of the inner B-field or catastrophic 
instabilities in the magnetosphere~\cite{ThompsonD01, Lyutikov03}.

Several arguments, besides spin-down via magnetic dipole radiation, 
indicate external field strengths  of $B_{\rm d} \sim 10^{14} - 10^{15} $~G for 
the majority of presently-known AXPs and SGRs\cite{ThompsonD95, VietriEA07}. 
In a few class members, as well as in some radio pulsars 
displaying magnetar-like bursting activity, sub-critical values $B_{\rm d} \sim 10^{13}$~G 
have been inferred, which are close to the upper end of the distribution in 
rotationally powered neutron stars \cite{ReaEA10, ReaEA12}. 
The strength of the interior magnetic field is estimated through energy arguments. The B-field energy is 
$E_{\rm B,int} \simeq (B_{\rm int}^2/8\pi) (4\pi R_*^3/3) \sim 2 \times10^{47} B_{\rm int, 15}^2R_{*,6}^2$~erg 
with $R_*$ the neutron star radius. All subscripts in this chapter indicate power of 10 values in 
CGS units,  {\it e.g.}  $B_{15} \equiv B/(10^{15}$G$)$ and $R_6\equiv R/(10^6$cm$)$,  
unless specified otherwise. $E_{\rm B,int}$ must be high enough to 
power the emission of SGRs and AXPs over their lifetime, as estimated from 
the spin-down age and/or the age of the supernova remnants  they are associated to. 
The persistent emission of the brightest members of the two classes 
($\sim10^{34}-10^{35}$~erg/s) provides 
an interior field estimate of B$_{\rm int} \sim 10^{15}$~G. 
The energy release and estimated recurrence rate of the initial
sub-second spike of the 2004 Dec 27 giant flare from SGR1806-20 
implies an energy release in the  $\sim 10^{49}$~erg range, 
which converts to B$_{\rm int} \approx 10^{16}$~G \cite{StellaEA05}. 
Additional studies indicate that magnetars contain interior B-fields well in excess of their already strong exterior dipoles ({\it e.g.} \cite{ReaEA10, ReaEA12, Dal12, Tie13, MakEA14, Ro-Ca16}), and numerical work dedicated to understanding their characteristic burst activity reached similar conclusions ({\it e.g.} \cite{Turo11, PoPe11}), pointing to values of B$_{{\rm int}, 15} \sim 5-20 $.
AXPs and SGRs, {\it i.e.} {\it classical magnetars}, 
possess a rotational energy of $E_{\rm rot} \sim 10^{44}-10^{46}$ erg, in most cases much lower than their magnetic energy.

According to current estimates \cite{BeniaminiEA19} magnetars form at a comparable rate 
to that~of ordinary NSs. The collapse of massive stars is believed to be their main 
formation channel, naturally relating newborn magnetars to core-collapse supernovae (CCSNe; see Sec. \ref{sec:SNe}) and long gamma-ray bursts (GRBs) in the collapsar scenario (see Sec. \ref{sec:GRB}).
Based on the stellar population properties of their associated
open cluster, progenitor masses of two classical magnetars 
have been constrained to  $ \sim (30 - 45)$~M$_\odot$ in two cases
\cite{Muno06, Bib08} 
and $M \sim (15-19)$ M$_\odot$
in another case \cite{Dav09}. 

Very high values of the internal B-field are key to the 
magnetar model. Conservation of the progenitor's magnetic flux 
during core collapse, the so-called {\it fossil field} scenario, 
is expected to give rise to a maximum strength in the $\sim 10^{14}$ G range, 
and only in a small fraction of the cases (for a review see \cite{FerrarioEA15}).
Dynamo action in the proto-neutron star (PNS) seconds after its formation is required to generate stronger fields. This is likely driven by the interplay of differential rotation, which stretches the field in a predominantly toroidal configuration, and the onset of Tayler or magneto-rotational instabilities to close the dynamo loop \cite{DuncanT92, Spruit09, Ray20}. 
The resulting B-field amplification draws 
from the energy of differential rotation, which may amount to $\sim 10$\% 
of the rotational energy of the PNS, 
$E_{\rm rot}  = I\Omega^2/2 \approx 2 \times 10^{52} I_{45} P_{-3}^{-2}$~erg. Here $I$ is the moment of inertia, $\Omega$ the angular velocity and $P$ the spin period. 
The Tayler dynamo has been estimated to saturate at 
$\sim 1$\% of the energy density in differential rotation \cite{Braithwaite06}. 
The amplified B-field energy may thus represent a fraction $f_{-3}$ 
of the PNS rotational energy, implying   
that magnetars are born with much larger rotational than magnetic energy. It is their fast spindown, due to the~strong magnetic dipole~field, which makes them magnetically-dominated at a later age, $\sim 10^3-10^5$~yr, when they are observed as AXPs/SGRs.
Relating the inner B-field to the fraction $f$ of spin energy in the PNS that goes to magnetic energy 
 gives $B_{\rm int}  \approx 10^{16}~G f^{1/2}_{-3} I^{1/2}_{45} P_{-3}^{-1} R^{-3/2}_6$. An interior B-field $\sim 10^{16}$~G would thus require a magnetar spin period of $\sim (1-2)$~ms at birth\footnote{Note that the virial limit of $\lesssim 10^{17}$ G holds for NS interior B-fields (e.g. \cite{Rei09}).}. Correspondingly, an angular momentum of order 
$\sim 10^{48} - 10^{49}$~g\,cm$^2$/s must be retained in the progenitors' core pre-collapse 
or transferred by fallback accretion to the PNS.  

That a stable, ms-spinning magnetar may form in a binary NS (BNS) merger~was first shown in numerical simulations by \cite{GiacPer13} and further studied, {\it e.g.} \cite{Shi17, Bai17, Hind18} especially in the aftermath of the 
BNS merger GW 170817. The large orbital angular momentum of the two NSs at the time of coalescence guarantees a remnant's~ms-spin period and rotational energy of $\gtrsim 3 \times 10^{52}$~erg.  Also in this case differential rotation leads to 
the fast growth of a $\sim 10^{16}$ G, mostly azimuthal B-field inside the merger remnant.In \cite{Ober17} it was found that, in the early phases, the large-scale dipolar B-field ($B_{\rm d}$) contains only a 
fraction of the poloidal magnetic field energy, most of it being concentrated in smaller-scale, higher-order multipoles. The 
possibility to form stable magnetars in BNS mergers  
depends sensitively on the not well constrained maximum mass of a NS, which is set 
by the equation of state of ultradense matter. 
For component masses of $\sim 1.3$~M$_\odot$, most NS mergers may end up producing~a~BH, possibly after the formation of a short-lived supramassive or hypermassive NS, temporarily sustained against collapse by centrifugal forces (due to rigid or differential rotation, respectively). However, if the maximum NS mass is $\gtrsim (2.3-2.4)$ M$_\odot$, the formation of a stable, massive NS is possible in a sizeable fraction of cases ({\it e.g.} \cite{Dal15, Pir17}).

Other channels
may  
lead to the formation of millisecond magnetars: besides BNS mergers, 
accretion induced-collapse of a white dwarf, merging of two white dwarfs or 
a white dwarf~and a NS ({\it e.g.} \cite{ZhongD20} and references 
therein).
Recent reviews on classical magnetars are \cite{MereghettiEA15, TurollaEA15, KaspiB17, EspositoEA18}. 

The bulk of a newborn magnetar rotational energy can be radiated away with enormous 
power in two main ways ({\it e.g.} \cite{OstrikerG69}): 

{\rm (a)} in the electromagnetic (EM) channel via magnetic dipole radiation, thanks to their intense (external) dipole B-fields. The EM luminosity is\footnote{This holds when the magnetic dipole is aligned with the spin axis. In general, if they are misaligned by an angle $\chi$, the rhs of Eq. \ref{eq:emspindown} must be multiplied by $\left(1+\sin^2 \chi\right)$ \cite{Spit06}.} 
\be
\label{eq:emspindown}
{L}_{\rm EM} (t) = \displaystyle \frac{\mu^2_{\rm d} \Omega^4(t)}{c^3}  =  \displaystyle  \frac{{E}_{\rm rot}}{\tau_{\rm EM}} \left(1+ \displaystyle \frac{t}{\tau_{\rm EM}}\right)^{-2} \approx 10^{47} \displaystyle \frac{{ E}_{{ \rm rot}, 52}}{\tau_{{\rm EM}, 5}} \left(1 + \displaystyle  \frac{t}{\tau_{\rm EM}}\right)^{-2} ~ {\rm erg}~{\rm s}^{-1} 
\ee
where $\tau_{\rm EM}$ is the characteristic time in which half of the initial spin energy is lost
\be 
\label{eq:temdef}
\tau_{\rm EM}  = E_{\rm rot}/L_{\rm EM} (t=0) \approx 1.4 \times 10^5 P^2_{-3} B^{-2}_{{\rm  d,} 14} I_{45} R^{-6}_6~~{\rm s} \, .
\ee
{\rm (b)} in the gravitational wave (GW) channel, {\it {\it e.g.}} as a result of the ellipticity $\epsilon_{\rm B}$ (and thus the 
mass quadrupole moment $Q \sim \epsilon_{\rm B} I$) induced by their ultra-strong (internal) magnetic field. The GW luminosity is\footnote{This holds when the magnetic symmetry axis is orthogonal to the rotation axis. In general, when they are misaligned by a tilt angle $\chi$, the rhs of Eq. \ref{eq:gwspindown} must be multiplied by $\left(1+ 15 \sin^2 \chi\right)/16$.}
\begin{eqnarray}
\label{eq:gwspindown}
L_{\rm GW} (t) & = & 32 G Q^2\Omega^6(t)/5c^5 =  \nonumber \\
& = & \displaystyle \frac{E_{\rm rot}}{\tau_{\rm GW}} \left(1+ \displaystyle\frac{2t}{\tau_{GW}}\right)^{-3/2}\approx 10^{48} \frac{{E}_{{\rm rot}, 52}}{\tau_{{\rm GW}, 4}} \left(1 + \displaystyle \frac{2t}{\tau_{\rm GW}}\right)^{-3/2} {\rm erg~s}^{-1} \, .
\end{eqnarray}
 The magnetically-induced ellipticity (see Sec. \ref{sec:GW}) is $\epsilon_{\rm B} \sim 4 \times 10^{-4} B^2_{{\rm int}, 16} R^4_6 M^{-2}_{1.4}$, $M_{1.4}$ being the PNS mass in units of 1.4 M$_\odot$. The characteristic spin-down time is 
\be
\tau_{\rm GW} = E_{\rm rot}/L_{\rm GW}(t=0) \approx 1.8 \times 10^4 P_{-3}^4 B^{-4}_{{\rm int,} 16} M^4_{1.4} I^{-1}_{45} R^{-8}_6 ~~{\rm s} \, .
\ee

\noindent
Magnetars are thus believed to be rotationally powered in their early infancy and 
magnetically-powered ({\it i.e. classical magnetars}) in their youth. 
\begin{figure}[ht]
\centerline{\includegraphics[scale=0.22]{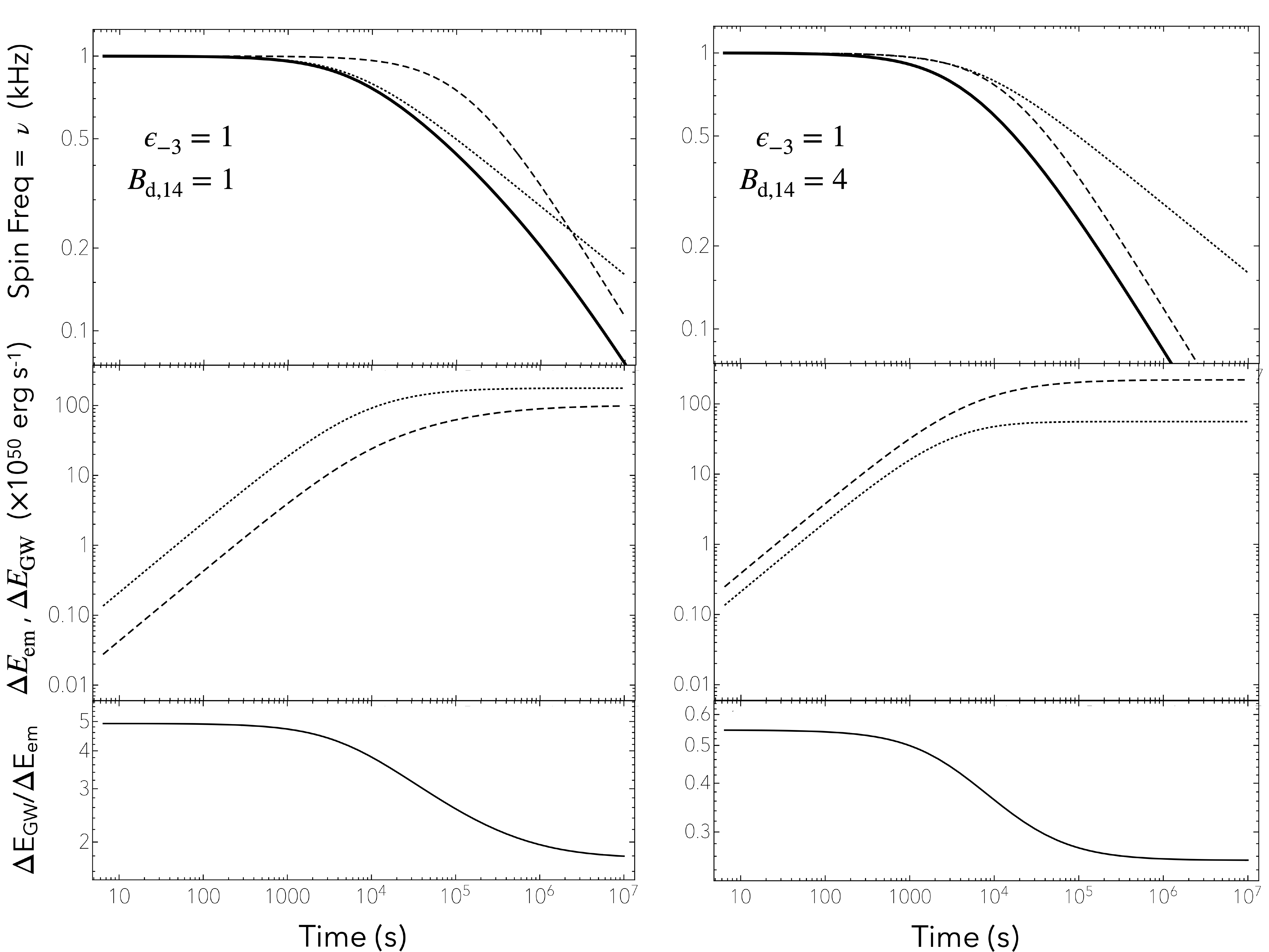}}
\caption{Spindown of 1-ms spin NS subject to magnetic dipole and GW losses. The two panels represent 
a GW-dominated (left) and EM-dominated (right) spindown, respectively. In both cases, the upper row depicts the spin frequency evolution (continuous), with the dotted (dashed) curve showing the spin evolution if only GW (EM) losses were in place. The middle row shows the total spin energy lost to GWs (dotted) and EM radiation (dashed) vs. time ({\it i.e.} the two luminosities integrated up to that  time). The lower row reports the ratio of the radiated energies vs. time.}
\label{fig:sd}
\end{figure}
Newborn, millisecond spinning magnetars have been invoked as the power source for a variety
of short-lived paroxysmal phenomena, such as 
GRBs (Sec. \ref{sec:GRB}), SNe~
(Sec. \ref{sec:SNe}),  GW events (Sec. \ref{sec:GW}) and Fast Radio Bursts (FRBs, Sec. \ref{sec:FRB}), by virtue of~the very fast rate 
at which they can tap and radiate away their rotational energy\footnote{They may release magnetic energy as well, though this channel involves a smaller reservoir
and there is no simple prescription for the power that can be liberated; this possibility has been discussed in relation to 
FRBs and the low-luminosity end of the short-GRB population.}.

\section{Millisecond magnetars as Gamma-Ray Burst central engines}
\label{sec:GRB}
GRBs are among the most powerful cosmic explosions, releasing $E_0 \sim 10^{49}-10^{52}$~erg, once corrected for a typical beaming $f_{\rm b} \sim 10^2$, in a 
relativistic jet of photons, pairs, magnetic fields and a small amount of baryons \cite{Pac86, Good86, She90, Mes93, Pi93} (see \cite{KuZha15} for a recent review). The jet moves with bulk Lorentz factor $\Gamma \gtrsim 10^2$, sending a relativistic shock-wave through the circumstellar and/or interstellar medium. 

Photons from the relativistic jet, Doppler-boosted to $\gamma$-rays (up to $\gtrsim 100$ MeV, or even $\gtrsim$ GeV  in some cases; e.g. \cite{FER19}), form the GRB {\it prompt} phase. Based on their duration, GRBs are classified as long ($>2$ s) or short ($<2$ s; \cite{Kou93}). This classification 
captures fundamental differences between the two GRB groups, including (i) their characteristic luminosities, long GRBs typically being $\sim$ 1-2 decades brighter 
and more energetic than short GRBs,  
(ii) their spectral properties, with long GRBs presenting a softer low-energy part of the spectrum (while the distributions in peak energy are very similar, with $E_{\rm p} \sim 10^2-10^3$ keV; \cite{Ghirla04, Ghirla09}), and (iii) their progenitors and host galaxies. 

Long GRBs are usually associated to a young stellar population as indicated by their often irregular, star-forming galaxy hosts and, within these hosts, to regions with the highest star-formation rate \cite{Fru06}. Some long GRBs were found to be associated with powerful type Ic SN ({\it e.g.} \cite{Gal98}, \cite{Hj03}), a class of core-collapse supernovae lacking H and He lines and thus originating from progenitors whose H- and He-rich envelopes were stripped before collapse (see Sec. \ref{sec:SNe}). The comparatively small size of the progenitor at the time of collapse thus facilitates the break-out of a GRB jet from the star \cite{Woo93,Matz03,Woo11}. This suggested, in particular, a connection with WR stars\footnote{
We note in passing that the two classical magnetars for which a progenitor mass of $ \sim (30 - 45)$~M$_\odot$  
has been derived are associated to young open clusters ($<8$ Myr) hosting WR stars.}
\cite{Woo93, McFWoo99}, which have $M > 20$ M$_\odot$ at birth ({\it e.g.} \cite{Cro07}) and suffer large mass losses during their evolution. At the time of collapse, they may have $M \gtrsim 10$ M$_\odot$ and $R \sim (1-10)$ R$_\odot$, as observed in the Milky Way ({\it e.g.} \cite{Zha19}). 

Concerning short GRBs, at least a fraction of them  
are produced by BNS mergers (and, possibly, BH-NS systems with not too large mass ratio), which  track an~old stellar population and are typically found in the outskirts of their massive host galaxies. GW events provide 
an estimated rate of $\sim 10^3$ Gpc$^{-3}$ yr$^{-1}$ for BNS mergers~\cite{Abb19}.  
The short-GRB rate at L$_{\gamma, {\rm iso}} > 10^{50}$~erg/s is $\sim (0.5-3)$ Gpc$^{-3}$ yr$^{-1}$ \cite{WaPi15, GhiEA16, GhiEA19}. The two rates can be reconciled if GRBs jets have a typical beaming $\sim$ 500, or a jet core size $\theta_c \sim 3.5^\circ$: this matches nicely the estimated jet core size in GRB 170817A (see below) \cite{Moo18a, Moo18b, Troj18, GhiEA19}. 

The characteristic energies and  fast variability time scales of prompt emission (down to $\Delta t \sim 1$~ms) suggest that GRBs originate in catastrophic events, involving the binding energy $W \sim GM^2/R \sim 10^{53} \left(M/M_\odot\right)^2 R^{-1}_6$ erg of a stellar mass compact object, or its somewhat lower rotational energy. Two possible central engines fulfilling these broad requirements for prompt emission have been considered: \\
(a) Hyper-accreting black holes (BHs) surrounded by an ultra-high density, thick~accretion disk of high angular momentum fallback material from the progenitor~star \cite{Woo93, Pop99,Nar01}. The innermost stable circular obit in the Schwarzschild potential requires that the specific angular momentum $\ell > \sqrt{12}GM/c = \sqrt{3} r_s c \sim 5 \times 10^{16} \left(M/ 3 M_\odot\right)$ cm$^2$ s$^{-1}$ ({\it e.g.} \cite{Shap83}), where $r_s$ is the Schwarzschild radius, placing a lower limit to $\ell$ 
in the pre-collapse stellar core for a disk to form. 
The jet-launching mechanism lies in the interaction of the inflowing disk plasma with the spinning BH, mediated by the plasma-frozen B-field which endows the BH with an effective magnetosphere as long as matter inflow continues  \cite{BlaZna77, MacTho82, Tche11, Tche20}. The accretion (viscous) timescale in a thick disk, $t_v \sim r^{3/2}/\alpha \sqrt{G M} h^2 \sim 0.1 - 20~(M/3 M_\odot)$~s for radii $\sim (3-100)~r_s$ and typical parameter values (viscosity $\alpha \sim 0.01$, thickness-to-radius ratio $h \sim 0.5$), in agreement with the duration of the prompt emission. \\ 
(b) Millisecond spinning, highly magnetised NSs, independently proposed as GRB central engines as far back as 1992 
\cite{Us92, DuncanT92}. Early suggestions focused on the release of their large magnetic energy reservoir, due to the short characteristic timescales of magnetic instabilities/reconnection events. 
If instead the (even larger) spin energy of a millisecond magnetar is responsible for the prompt GRB emission,
a more powerful mechanism than the classical dipole formula 
(Eqs. \ref{eq:emspindown}-\ref{eq:temdef}) is required for plausible 
values of $B_{\rm d}$ and $P$. It was then proposed that energy may be extracted very efficiently, in the first $\lesssim 30$ s of the NS life, by a strong neutrino-driven MHD wind from the NS surface\footnote{During this short time, the enhanced spin down luminosity can be  expressed as (\cite{Metz11}) 
\be
L_{\rm wind} = L_{\rm EM} \left(R_L/R_Y\right)^2 \times {\rm max}\left[\sigma_0^{-/1/3}, 1\right]\, .
\ee
Here $R_L = c/\Omega$ is the light cylinder, $R_Y \leq R_L $ the equatorial radius of the closed magnetosphere and $\sigma_0 = \phi^2 \Omega^2/(\dot{M} c^3)$ the magnetization parameter of the wind, where $\dot{M}$ is the neutrino-driven mass loss from the NS surface and $\phi$ the magnetic flux threading the NS surface and linked to open B-field lines. In the early stages the closed magnetosphere is small, {\it i.e.} a large fraction of the NS magnetic flux is linked to open B-field lines, hence $R_Y \ll R_L$. In addition, the neutrino-driven wind is strongest, carrying a sufficiently large mass outflow to ensure a low magnetization  ($\sigma_0\ll 1$): as a result, $L_{\rm wind} \gg L_{\rm EM}$. Later, as the mass loss rate drops, $\sigma_0$ grows quickly and the wind approaches a force-free condition ($\sigma_0 \rightarrow \infty$) while the closed magnetosphere expands, $R_Y \rightarrow R_L$. Thus $L_{\rm wind} \rightarrow L_{\rm EM}$ and the classic magneto-dipole spindown kicks in.}, {\it i.e.} on a timescale much shorter than $\tau_{\rm EM}$ 
\cite{Metz08, Buc09, Metz11}. 
Observationally, the (beaming-corrected) energy needed to power even the brightest prompt emissions rarely exceeds $\sim 10^{52}$ erg. On the other hand, the maximum spin energy of a NS is $\sim (3-8)\times 10^{52}$ erg, depending on its mass and equation of state \cite{margametz17, Dal18}. Thus, a GRB prompt phase requires a fraction of the maximum~spin 
energy of a millisecond magnetar. Moreover, the specific angular momentum of the latter is $\ell = \sqrt{2 I E_{\rm rot}}/M \approx 3 \times 10^{15} R^2_6 P_{-3}^{-1}$ cm$^2$s$^{-1}$, imposing a less demanding constraint on the angular momentum of the progenitor's core compared to an hyper-accreting BH.

In both long and short GRBs the measured prompt $\gamma$-ray emission may amount to a seizable fraction of the total energy release ($\gtrsim 50$\%; \cite{Fan06, Zhang07c}), challenging theoretical models and requiring
 an efficient mechanism to convert the jet power to high-energy radiation. Proposed mechanisms for the prompt emission include internal shocks generated by collision of shells traveling at different values of  $\Gamma$ ({\it e.g.} \cite{ReeMes94, DaMok98}) or by internal magnetic dissipation ({\it e.g.} \cite{DreSpr02, LyuBla03, ZhY11}), with a possible contribution from bulk Comptonisation of photospheric emission ({\it e.g.} \cite{Ry05, Bel10, Gui13, Vur13})). 
 Due to the observed fast variability, causality requires that the size of the prompt-emitting region be $\lesssim c \Delta t \Gamma^2 \sim 3 \times 10^{11} \Delta t_{-3} \Gamma^2_{100}$ cm.

GRB {\it afterglows} are observed from X-ray to radio wavelengths for timescales of weeks/months in some cases and originate from the relativistic shock-wave resulting from the jet interaction with the surrounding medium \cite{PaRh93}, \cite{ReesMesz97}, \cite{SaPir98}.
Afterglows are produced farther out from the central engine, at a distance $> 10^{16}$ cm, where, having swept up a mass $M_{\rm dec} \sim {E}_0/\left(\Gamma^2 c^2\right) \sim 10^{-7} {M}_\odot ~{E}_{0,51} (\Gamma/100)^{-2}$, the {\it external shock} decelerates and converts its kinetic energy to the energy of relativistic particles that radiate it away.

At the end of the prompt phase, the X-ray lightcurves of most long GRBs show a short-lived, steep flux decay $\propto t^{-3}$ (or steeper), 
followed by a shallow decay phase, or {\it plateau}, which is always flatter than $t^{-0.8}$ (typical index $\sim -0.5$)  and lasts up to $t \gtrsim 10^4$ s \cite{Nous06}. Short GRBs display a similar phenomenology, albeit less frequently and on typically $\sim$ 10 times shorter timescales. Plateaus usually end with a smooth transition to the so-called normal decay law (slightly steeper than $\sim t^{-1}$) that characterises afterglows, a behaviour expected from the decelerating shock-wave.

The occurrence of the X-ray plateau cannot be explained by invoking spectral transitions in the emitting region, because observations indicate a constant X-ray spectrum throughout the whole duration of the plateau. Since X-ray plateaus are observed in long as well as short GRBs, any explanation for this feature should be general enough to hold in both classes of events.A straightforward interpretation involves the presence of a long-lived central engine, releasing power at a nearly constant rate for the typical duration ($\lesssim 10^4$ s in long GRBs and $\lesssim 10^3$ s in short GRBs; \cite{Zha19}). The energy may be injected in the afterglow-producing external shock, {\it e.g.} through a (magnetized) wind, constantly refreshing its kinetic energy and thus maintaining a constant level of radiation \cite{Dai98, ZhaMe01, Dal11, Gomp15}. Alternatively, the power from the central engine may give rise to an additional emission component temporarily outshining 
the external shock emission ({\it e.g.} \cite{Lyons10, Rowl13, SieCio16}).

The  
plateau durations in long GRBs, $\sim 10^3-10^4$ s or longer in some cases, place a strong constraint on the central engine activity. Models with a central BH were devised in which the density profile in the progenitor star is tuned to provide the long-term time-varying mass accretion rate through the ultradense disk, as required to account for the X-ray afterglow light curve shape ({\it e.g.} \cite{Ku08}). 
 By contrast, a millisecond spinning magnetar would have just the right spin down luminosity and plateau duration (Eq. \ref{eq:emspindown}), thus offering a straightforward interpretation. This idea was put forward \cite{Dai98, ZhaMe01} well before the Swift discovery of X-ray plateaus. By using the magnetic dipole spindown formulae (Eq. \ref{eq:emspindown}), \cite{Lyons10} and \cite{Rowl13} obtained fits of X-ray plateaus in short GRBs deriving birth spin periods $P\sim (0.5-10)$ ms and dipole magnetic fields $B_{\rm d}\sim (10^{14} - 10^{16}$) G. 

In \cite{Dal11} the luminosity evolution of a relativistic external shock was calculated based on a more advanced model including (i) energy injection from the magnetar spin down, (ii) radiative losses with efficiency as a free parameter, (c) a simple hydrodynamic prescription for the evolution of $\Gamma$ with (observer's) time. The solution to this model was used to fit 4 long GRB plateaus,  deriving $B_{\rm d} \sim (10^{14} - 10^{15})$ G and $P\sim (1-5)$ ms as birth parameters. Application to a sample of 64 long GRBs confirmed previous results and extended the birth parameter ranges \cite{Berna12}. This work also highlighted  the existence of a correlation between the plateau luminosity and its duration, further studied and characterised in \cite{Daino13}. 
The model was further extended with a more general expression for the magnetar spindown \cite{Stratta18}, and used to fit the light curves of 51 long and short GRBs with X-ray plateaus, obtaining  $P\sim (0.8-20)$ ms and $B_{\rm d} \sim (0.5-20) \times 10^{15}$ G (cf. \cite{Li18}). A correlation between $P-$ and $B_{\rm d}-$values, with short GRBs lying systematically on the long-spin and strong-$B_{\rm d}$ end of the distribution while long GRBs occupy the opposite  extreme. 
\begin{figure}[ht]
\centerline{\includegraphics[scale=0.215]{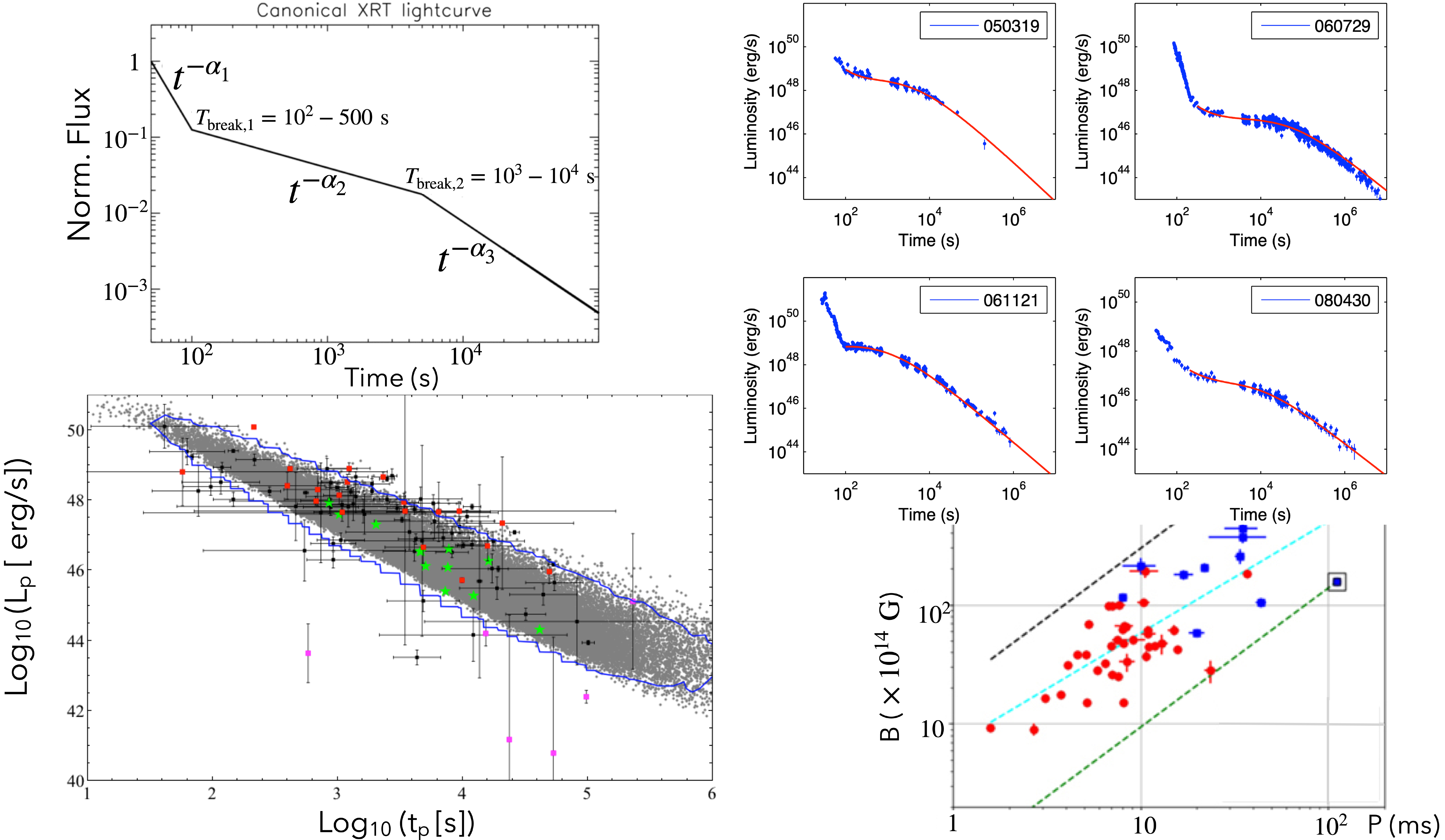}}
\caption{{\it Upper Left:} sketch of the canonical GRB X-ray light-curve (adapted from~\cite{Corsi09}). The prompt phase ends with a steep flux decay ($\alpha_1 > 3$), which is followed by a shallow decay phase ($0 \leq \alpha_2 < 0.8$; ``plateau") accompanied by a spectral hardening. The plateau joins smoothly to the so-called ``normal" decay, {\it i.e.} the one expected from standard afterglow theory; {\it  Upper right:} plateau light-curve fits of four selected GRBs with an energy injection model invoking the spindown luminosity of a millisecond magnetar. Radiative losses are also included parametrically (from \cite{Dal11}); {\it Lower left:} the anti-correlation between the luminosity at the end of the plateau ($L_p$) and the plateau duration ($t_p$) for a large sample of long GRBs (from \cite{Berna12}); {\it Lower right:} the $B_d$ vs. $P$ correlation for a sample of long {\it and} short GRB plateaus that were fitted with an energy injection model invoking a spinning down millisecond magnetar (from \cite{Stratta18}). It was noted that this correlation matches the well studied  ``spin-equilibrium" line for accreting NS in galactic binaries, once the mass accretion rate is normalised to typical values expected in a GRB prompt phase ($\dot{M} \sim 10^{-4}-0.1~{\rm M}_\odot$ s$^{-1}$).}
\label{fig:plateau}
\end{figure}

A test of the energy injection model with magnetar spin-down vs. a fireball model with no energy injection in two particular short-GRBs concluded that the data favor the former model with great significance, requiring NS matter to have a relatively stiff equation of state, with a maximum non-rotating mass of $\approx 2.3$ M$_\odot$ \cite{Sa19}.

An open problem for the magnetar scenario is the great diversity observed in the joint optical/X-ray light curves of GRB plateaus ({\it e.g.} \cite{Pana06, Li07, Gomp15}). Approximately half of GRB afterglows appear to display an achromatic behaviour \cite{KuZha15}, in which the optical and X-ray emission have a similar temporal evolution as expected if they were both produced by the forward shock. However, in other afterglows these two components evolve differently \cite{KuZha15}, even though the X-ray light curve, from the start of the plateau through the transition to normal decay, typically does not show any spectral evolution. The X-ray behaviour would thus suggest that the transitions are still due to geometrical and/or hydrodynamical factors, but the optical behaviour argues against this possibility. This points to the existence of two separate emission regions at least in a fraction of GRBs (see, however, \cite{Gomp15}). In the magnetar scenario, this may occur if the NS spin energy is not injected in the external shock but rather gives rise to a magnetar wind nebula (MWN) as envisaged in (\cite{SieCio16}; cf. \cite{MePi14, MargetAl18}). In this case, the X-ray plateau would reflect the emission from the MWN, while the optical would be dominated by emission from the forward shock. 
This may be possible in short GRBs, whereas in long GRBs the observed X-rays would hardly be able to diffuse through the 
several M$_\odot$ of ejecta in a $\sim 10^4$ s timescales.

The discovery of the gravitational signal GW~170817 and of the associated short GRB 170817A ({\it e.g.} \cite{Abb17}) confirmed the BNS merger nature of at least some short GRB progenitors. Broadband afterglow observations showed that a relativistic structured jet \cite{MesRee01, Ro02} was launched right after the merger 
\cite{Moo18a, GhiEA19}, having a wide angular profile of 
 kinetic energy with a narrow core ($\theta_c \sim 5^\circ$) and an extended decaying tail at larger angles \cite{Margu18, Troj18, Moo18b}. Following this discovery it was proposed \cite{Oga20} that the X-ray plateau of GRBs can be 
 interpreted as a consequence of the jet's extended angular profile - combined with its relativistic motion - when observed along the jet axis, requiring a relatively large emission radius, $R > 10^{14}-10^{15}$~cm.~
It~was further shown that steep decays and plateaus that follow the prompt are generic features of structured jets even if observed off-axis, although their slopes and durations depend sensitively on the viewing angle \cite{Asc20}. In this alternative scenario the plateau carries signatures of the jet structure, but not of the 
central engine. Therefore all considerations about a magnetar central engine (but the plateau) remain unaltered.
\section{Millisecond magnetars and Supernovae} 
\label{sec:SNe}
A group of stellar explosions exceeding by $1-2$ orders of magnitude the peak luminosity 
of standard SNe has been identified and studied in the last two decades. These events, 
termed SuperLuminous SNe (SLSNe), occur $\sim 100$ times less frequently than SNe and 
are broadly classified into hydrogen-poor (SLSNe - I) and hydrogen-rich (SLSNe - II) 
types (for reviews see \cite{Gal-Yam19, Inserra19}). They are believed to arise in the core collapse  
of massive stars ($\sim 20-40$ M$_\odot$). 
Different scenarios have been investigated to explain 
the extra energy required to power SLSNe and attain peak luminosities of 
$\sim$ few $\times\, (10^{44}- 10^{45})$~erg/s; among these the formation of a 
millisecond spinning magnetar appears to be especially promising, particularly in relation 
to SLSNe - I. 

Originally proposed to interpret the characteristics of a peculiar Type Ib SN (SN 2005bf, \cite{MaedaEA07}),
the magnetar model for SLSNe was further developed in \cite{KasenB10, Woosley10, DessartEA12}. The model envisages that the spin-down 
power of the newborn millisecond magnetar is emitted in the electromagnetic channel in the form of
magnetic dipole radiation and pulsar wind, and deposited deep in the expanding SN layers, from which 
it diffuses outwards. We summarise here the basic features of the simple model by
\cite{KasenB10}.
 
In the core-collapse of the progenitor star of radius $R_*$, a mass $M_{\rm ej}$ is ejected with
velocity $v_{\rm ej}$, such that the initial energy of the SN is $E_{\rm SN} = M_{\rm ej} v_{\rm ej}^2/2\sim 10^{51}$erg. 
In the absence of other energy sources (such as decay of $^{56}$Ni, fallback accretion or magnetar
spin-down), self-similar adiabatic evolution ensues with expansion time 
$t_{\rm e} = R_*/v_{\rm ej}$. The internal energy decreases as $E_{\rm int} \sim E_{\rm SN} (R_*/R)$, with 
$R = v_{\rm ej} t$ the radius of the remnant. When the age of the remnant equals
the radiative diffusion timescale $t_{\rm d} \sim (\kappa M_{\rm ej}/v_{\rm ej}c)^{1/2}$ (here $\kappa$ 
is the opacity) the light curve attains its maximum luminosity of 
$\sim E_{\rm SN} t_{\rm e}/t_{\rm d}^2 \lesssim 10^{43}$~erg/s. 

For plausible amounts of $^{56}$Ni the extra 
heat source provided by  
radioactive decay is not enough to reach 
maximum luminosities up to $\sim 10^{45}$~erg/s. 
Instead the rotational power of a 
newborn magnetar suffices. 
The bulk of its rotational energy $E_{\rm rot}$ is released on a spin-down 
timescale $\tau_{\rm EM}$ (Eq. 1) which is shorter than $t_{\rm d}$ if 
\begin{equation}
B_{{\rm d,} 14}P_{-2}^{ -1} > 2 M_{{\rm ej},5}^{-3/8} E_{51}^{1/8} \, , 
\end{equation}
here $M_{{\rm ej,}5} = M_{\rm ej}/5M_{\odot}$. 
 The rotational energy of the remnant will 
become dominant when $E_{\rm SN} (R_*/R) < E_{\rm rot}$ 
(since usually $(R_*/R) \ll 1$ this condition does not require extreme  
values of $E_{\rm rot}$). 
In this case the SN attains a peak luminosity  of 
\begin{equation}
 L_{\rm p} \sim E_{\rm rot} \tau_{\rm EM}/t_{\rm d}^2 \sim  5\times10^{43} B_{{\rm d,} 14}^{-2}M_{{\rm ej,}5}^{-3/2}E_{51}^{-1/2}
{\rm erg/s}\,   
\end{equation}
The above model provides only rough estimates, which nevertheless
capture the characteristic scales of the magnetar scenario. 
In their more advanced model \cite{KasenB10} 
explore other effects such as those 
arising from magnetar energy release for $t > \tau_{\rm EM}$
and derive analytical expressions for the SN light curve
in the regimes arising from a different hierarchy of timescales. 

Magnetar models were applied to SLSNe light curve fitting 
by a number of authors. Based on the analysis of the 
bolometric light curve of 5 SLSNe-I, \cite{InserraEA15} estimated the following
ranges for the magnetar parameter  $P_{-3} \sim 1.7-7$, $B_{{\rm d,} 14} \sim 3-7$, $E_{51}\sim 0.4 - 7$ 
and an ejected mass of $M_{\rm ej} \sim 2.3 - 8.6$~M$_{\odot}$.
By fitting of the multicolor light curves of 38 SLSNe-I with a more general model 
\cite{NichollEA17} derived the following (1$\sigma$) parameter ranges: 
$P_{-3} \sim 1.2 - 4$, 
$B_{{\rm d,} 14} \sim 0.2 - 1.8$, $E_{51}\sim 2 - 10$ and $M_{\rm ej} \sim 2.2 - 13$~M$_{\odot}$.  Applications to hydrogen-rich SLSNe (SLSNe-II) have also been proposed \cite{OrellanaEA18}.

Millisecond magnetar models for SLSNe also face some difficulties: for instance early or 
late bumps in the light curves are not straightforwardly 
interpreted, and the observationally-determined ejected masses are sometimes $3-4$ times larger than predicted. It has been proposed that the shock breakout of a magnetar-inflated bubble, or of a GRB-related relativistic jet, crossing the SN ejecta can produce an early-time secondary maximum in the SN light-curve as a specific signature \cite{KasMeBi16, MeETAL18}. 1D models and simulations are probably inadequate to capture such complex features in light curves and spectra. Multidimensional simulations of magnetar winds embedded in young remnants offer a better perspective and the first 3D studies have recently become available (see {\it e.g.} \cite{ChenEA20}). 
 
Accretion of fallback matter onto a newly formed fast spinning magnetar shortly after 
collapse may also contribute determining the magnetar properties as well as the energy 
release into the exploding SN envelope. In
\cite{PiroO11} the  
fallback matter inflow rate is parametrised with an initial growth (scaling as $\propto t^{1/2}$) followed 
a few hundred seconds later by a decrease, with the characteristic time dependence of fallback 
($\propto t^{-5/3}$).
If accretion of fallback matter onto the proto-magnetar does take place, 
the total amount of accreted matter must be small enough (several tenths of solar mass) 
that the maximum NS mass is not exceeded and collapse to a BH does not take place. 
This in turn requires that the initial spin  is in the millisecond range, $B_{\rm d} \geq 10^{15}$~G and fallback does not involve  
large masses. Under these circumstances at the 
end of the accretion phase, when the NS magnetospheric radius 
\begin{equation}
r_{\rm m} \simeq 10^6 B_{{\rm d,} 15}^{4/7} M_{1.4}^{-1/7} \dot M_{31}^{-2/7} \ {\rm cm} 
\end{equation}
becomes larger than the corotation radius 
\begin{equation}
r_{\rm c} \simeq 1.7\times 10^6 M_{1.4}^{1/3} P_{-3}^{2/3} \ {\rm cm} ,
\end{equation}
fallback matter will be flung outwards at the expense of the proto-magnetar 
rotational energy by the so-called {\it propeller} mechanism. 
Collision of this matter with the ejecta will shock-heat, creating a cavity deep inside the expanding envelope and depositing 
$\sim 10^{51}-10^{52}$~erg in the SN in 
$\sim$ tens of seconds. 
At the end of the propeller phase energy losses will be dominated by 
(slower-decreasing) dipole spin-down. 
The 
large energy release 
by the propeller in the early stages adds up to the 
whole budget of the SN, giving rise to a fast and powerful 
evolution.
In this scenario the energy injection 
takes place at later times, at variance with standard models for SLSNe in which the 
magnetar dipole spin-down losses directly power the observed light-curve.  

In \cite{MetzgerEA18} an extensive study of fallback in  millisecond magnetar models for 
both GRBs and SLSNe is presented in which additional regimes and effects are explored. Among these it is considered
that fallback accretion might spin up an initially non-maximally rotating 
protomagnetar, at the same time enhancing the spin-down luminosity with respect to
the standard dipole formula (see footnote 5). 
This effect is found to reduce the gap in the range of initial spin periods and B-fields 
required to model GRBs and SLSNe with millisecond magnetar models.

An open issue with the above scenarios (and, more generally, core collapse formation 
of millisecond magnetars) is that the Supernova Remnants (SNRs) originating in stellar explosions 
which receive an additional energy of $\sim 10^{51} - 10^{52}$~erg from magnetar spin-down
should be distinctly different from shell-type SNRs hosting lower magnetic field NSs. 
This is contradicted by the SNRs associated to classical magnetars, which display 
fairly standard features \cite{VinkK06},\cite{MartinEA14}. 
Possible way outs have been 
proposed. A very fast expanding envelope may dissipate all its energy in several hundred years and make the SNRs too faint to be detected: in fact more than half presently-known magnetars are not associated to SNRs. Alternatively the bulk of the rotational energy of millisecond magnetar could be released in the GW channel, thus leaving the expanding remnant unaffected (\cite{Dal07}; see Sect. \ref{sec:GW}).

\section{Gravitational waves from millisecond spinning magnetars}  \label{sec:GW}
The huge rotational energy of millisecond magnetars may also give rise to powerful GW emission if the NS develops a sufficiently large, time-varying mass quadrupole moment (Q). Non axisymmetric shape distortions, caused 
{\it e.g.}  by a superstrong interior B-field or by  
an extreme rotation rate, can produce this effect.
The interior B-field generally induces an ellipsoidal deformation of a NS shape. 
If the spin axis of the ellipsoidal NS is {\it not aligned} with one of the principal axes, the NS will undergo free-body precession. In these conditions,   
a viscosity-driven  secular instability\footnote{Secular instabilities arise from the fact that 
 lower-energy states are accessible to the fluid if it can get rid of its excess energy, {\it e.g.} via GW-emission or viscosity on timescales $\gg$ the dynamical time.} 
 can operate 
 \cite{Jo76, Cut02}:  viscosity will dissipate spin (precession) energy while conserving angular momentum ($L$), hence the NS will end up spinning, after a few viscous times ($\tau_{\rm visc}$), around the axis with the largest moment of  inertia\footnote{Indeed, for a constant $L = I \Omega$, the spin energy $T = L^2/ 2I$ is minimised by maximising $I$.}. 
  
Let us approximate for clarity the NS as a biaxial ellipsoid with $I_x = I_y~\neq~I_z$, where ($x, y, z$) are the directions of the three principal axes and $z$ represents the symmetry axis. Oblate ellipsoids are characterised by $I_z > I_x$: thus, viscous dissipation will lead them to spin around the $z$-axis, with an axisymmetric shape in the orthogonal plane ($I_x = I_y$). Prolate ellipsoids have the opposite property, $I_x > I_z$, and therefore will spin-flip and end up rotating, after a time $\sim \tau_{\rm visc}$, around the $x$-axis with the non-axisymmetric shape in the orthogonal plane  (the symmetry axis is orthogonal to the spin axis). 
The former case would result in $Q= 0$, hence no GW emission. The latter case instead maximises $Q \sim \epsilon I_x$, where the ellipticity $\epsilon = \left(I_x - I_z \right) / I_x$, thus making the NS  an efficient GW emitter under suitable 
circumstances according to Eq. \ref{eq:gwspindown} \cite{Cut02, StellaEA05,Dal07, Dal09, Dal15, Dal18, LaJo17, LaJo20}.
 
In the dynamo scenario, a toroidal component is expected to dominate the interior B-field of magnetars  
(\cite{ThompsonD93, ThompsonD96, Spruit09, GiacPer13}). The azimuthal magnetic field, in turn, induces a prolate shape distortion of the NS, with ellipticity $\epsilon_{\rm B}$ and with the magnetic axis as its symmetry axis ({\it e.g.} \cite{Cut02} and references therein). Thus, newborn magnetars are ideal candidates for the spin-flip instability which, combined with the ms spin, makes them optimal GW sources among newborn NS. 

The magnetically-induced ellipticity is determined by both the strength and geometry of the interior B-field ({\it e.g.}, \cite{MaEA11, MaEA13, Hask08, Ak13, CioRe13, Dal15}). Dimensionally $\epsilon_{\rm B} \sim E_{\rm B}/W$, where the NS magnetic energy $E_{\rm B} \sim 2 \times 10^{49}B^2_{{\rm int}, 16} R^3_6$~erg. Models of the magnetic field distribution in  NSs 
consider mixed toroidal/poloidal components, like {\it e.g.} 
the so-called {\it twisted-torus} configuration 
in which a toroidal field~($B_\phi$) is confined 
within the NS core, 
in a torus-shaped region pierced by poloidal field~lines, ($B_{\rm p}$).  The dipole component of B$_{\rm p}$ forms the exterior dipole B-field, B$_{\rm d}$. 
Stability arguments \cite{Braithwaite06, Br09} require the toroidal-to-poloidal magnetic energy ratio $E_{{\rm B,} \phi}/E_{{\rm B,} p} \gtrsim 0.5$, while the 
maximum 
value for this ratio is less constraining and 
can be $\sim 10^2-10^3$, depending on the structure and size of the torus-shaped region ({\it e.g.} \cite{Ak13, Dal15, Glasky16}). Thus, $B_\phi$ can largely dominate the interior magnetic energy and the corresponding $\epsilon_B$-value. In this case we may write 
\be
\label{eq:epsilon}
\epsilon_{\rm B} = k\, (E_{{\rm B,} \phi}/W) \approx 4\times 10^{-4}  \left(k/4\right) B^2_{{\rm int,}16} R^4_6 M^{-2}_{1.4}
\, ,
\ee
where 
the geometrical factor $k$ is $\approx 4$ for a uniform B-field \cite{Cut02} and can be a few times larger in twisted-torus configurations ({\it e.g.} \cite{Mastra11, Ak13, Mastra13, Dal15}).

The spin-flip instability is generic to all rotating stars with misaligned toroidal B-fields \cite{MeTa72, Jon76, Cut02}. However, since L$_{\rm GW} \propto \epsilon_{\rm B}^2 \Omega^6$ and L$_{\rm EM} \propto B^2_{\rm d} \Omega^4$, it can~lead to strong GW signals only in millisecond spinning NS provided that 
(i) the magnetically-induced ellipticity is $\epsilon_{\rm B} \gtrsim 5 \times 10^{-4}$, requiring an interior magnetic field B$_{\rm int} \gtrsim 10^{16}$~G, (ii) the exterior magnetic dipole field is B$_{\rm d} \lesssim$ a few $\times 10^{14}$ G, in order to prevent magnetic dipole radiation from dominating the spindown and choking~the GW signal (in the latter case a bright EM transient may result, {\it e.g.} a GRB X-ray  plateau or a SLSN); (iii) viscous dissipation of free-body precession acts on a timescale $ \tau_{\rm visc} < \tau_{\rm EM}$, to allow GWs to kick in while the NS still retains its original spin\footnote{At  
these early times the NS temperature is T$\sim (3-10) \times 10^9$ K (\cite{Dal09}, \cite{Dal18}; \cite{LaJo20}), implying the main dissipative term should be bulk viscosity. The latter requires a periodic pressure perturbation in the fluid NS to activate out-of-equilibrium chemical reactions. The precessional motion of the fluid NS provides such a perturbation, the amplitude of which is still debated  
({\it e.g.} \cite{Dal18, LaJo17, LaJo20}).}. 

The GW signal emitted by the millisecond magnetar, once its symmetry axis has become perpendicular to the spin axis (orthogonal rotator), has frequency~
$f_{\rm GW}=2/P = \Omega/\pi$ and a relatively simple shape: a sinusoidal oscillation with decreasing amplitude ($h \propto \Omega^2$) and frequency ($f_{\rm GW} \propto \Omega$),  resembling in many respects a {\it time-reversed chirp} lasting $\sim$ hrs (\cite{Dal18}). The horizon distance for ideal matched-filter searches with Advanced LIGO/Virgo is estimated to be $\sim 20$ Mpc \cite{Cut02, StellaEA05, Dal09} within which one expects, conservatively\footnote{For this estimate, we adopt a minimum magnetar birth rate of one per 10 CCSNe.}, $\gtrsim 0.3$ magnetars per year to be formed \cite{Dal18}, given a local rate of  SN Ibc+II $\sim 7 \times 10^{-5}$ Mpc$^{-3}$ yr$^{-1}$ \cite{LiCho11}.

Very fast rotation leads also to a large mass quadrupole moment ($Q$) in fluid bodies, regardless of the B-field strength. The corresponding symmetry axis is aligned with the spin axis, hence no GW emission ensues. However
when the kinetic-to-binding energy ratio $T/ \left|W\right| 
\gtrsim 0.27$ (here $T=E_{\rm rot}$), the fluid becomes {\it dynamically} unstable \cite{Cha69},  transitioning to a new stable state on a timescale $\tau_{\rm dyn} \sim 2  \pi  \sqrt{R^3/\left(G M\right)} \lesssim 1$~ms. In the stable state the rotating body has a highly elongated, non-axisymmetric shape (bar-like) and $\lesssim 1$ ms spin, thus emitting GWs according to Eq. \ref{eq:gwspindown}, with  $\epsilon$ related to the degree of asymmetry of the bar.  The high energy threshold makes the dynamical instability unlikely to occur (\cite{Dimm08}; see, however, \cite{Shi05}).  Indeed, as the PNS is spun up by angular momentum conservation in the progenitor's collapsing core, it becomes subject to a {\it secular} bar-mode instability already at $T/ \left|W\right| \gtrsim 0.14$.  
This is the fastest growing ($f$-mode, $\ell =m=2, \tau_{\rm f} \sim 1-10^5$ s; \cite{LaiShap95}) among the so-called CFS (Chandrasekhar-Friedmann-Schutz) instabilities \cite{Cha70, Frie78}, that are secular GW-driven frame-dragging instabilities\footnote{As long as viscosity can be neglected.} of fast-rotating stars.  

In the case of the $f$-mode, the fluid becomes prone to a non-axisymmetric, elliptical deformation of its shape (Riemann-S ellipsoid)  with $\epsilon = (a^2_1-a^2_2)/(a^2_1+a^2_2)$, $a_1$ and $a_2$ being the semi-major axes of the elliptical cross-section orthogonal to~the spin axis. The elliptical pattern rotates, in the inertial frame, at the angular~frequency $\Omega_{\rm p}$, leading to GW emission which provides the required energy sink for the instability to grow, 
up to a large saturation amplitude ($\epsilon \lesssim 10^{-2}$; \cite{LaiShap95, Corsi09}). Superposed~to the pattern rotation is a field of internal motions with vorticity~$\zeta$ and angular frequency $\Lambda$ such that, in the inertial frame, the rotation of fluid particles has angular frequency\footnote{$\Lambda$ is the angular frequency of internal motions in the frame co-rotating with the elliptical pattern.} $\Omega = \Omega_{\rm p} + \Lambda$. At the onset of the instability, the energy lost to GWs drives the growth of the mode amplitude without affecting $\Omega_{\rm p}$~and~$\Lambda$ 
(the total circulation is conserved, see \cite{Corsi09}). Once the mode saturates 
GW losses start reducing $\Omega_{\rm p}$: the NS then follows a sequence of Riemann-S ellipsoids with decreasing pattern rotation and constant $\Lambda$, towards a final state which has interior fluid motions (vorticity $\zeta$) but a stationary (non-radiating) non-axisymmetric shape (Dedekind ellipsoid). We note that, up to this point, the model does not require the NS to have 
a magnetar-like B-field but only a millisecond spin.

As with spin-flip, magnetic dipole radiation due to the exterior B$_{\rm d}$ 
introduces an additional sink for the NS spin energy.  The latter is 
$E_{\rm rot} \gtrsim 5 \times 10^{52}$ erg if the secular instability~sets~in. 
It was shown in \cite{Corsi09} that this 
extra energy loss 
can (a) accelerate the instability growth rate and the NS evolution along the same sequence of Riemann-S ellipsoids and (b) extract a fraction of E$_{\rm rot}$ releasing it in the EM window. Because $\Omega$ (and $\Omega_{\rm p}$) is constant during the growth of the $f$-mode, a constant EM luminosity would result, $\sim 10^{47}$ erg s$^{-1}$ for typical magnetar parameters, which would account {\it e.g.} for an early GRB X-ray plateau. Like with the spin-flip instability, the NS spin down is initially only due to magnetic dipole radiation. However, in a time $\tau_{\rm f} \lesssim 10^3$~s the $f$-mode amplitude becomes large enough to make GW emission dominant. The early EM signal would thus be followed, after $\lesssim 10^3$ s, by a long-lived GW signal (lasting $\sim \tau_{\rm f}$) at the nearly constant frequency $f = 2 \nu_{\rm p} = \Omega_{\rm p}/\pi$. Both signals would then fade rapidly due to the strong NS spindown.

\begin{figure}[ht]
\centering
      \includegraphics[width=1.0\textwidth]{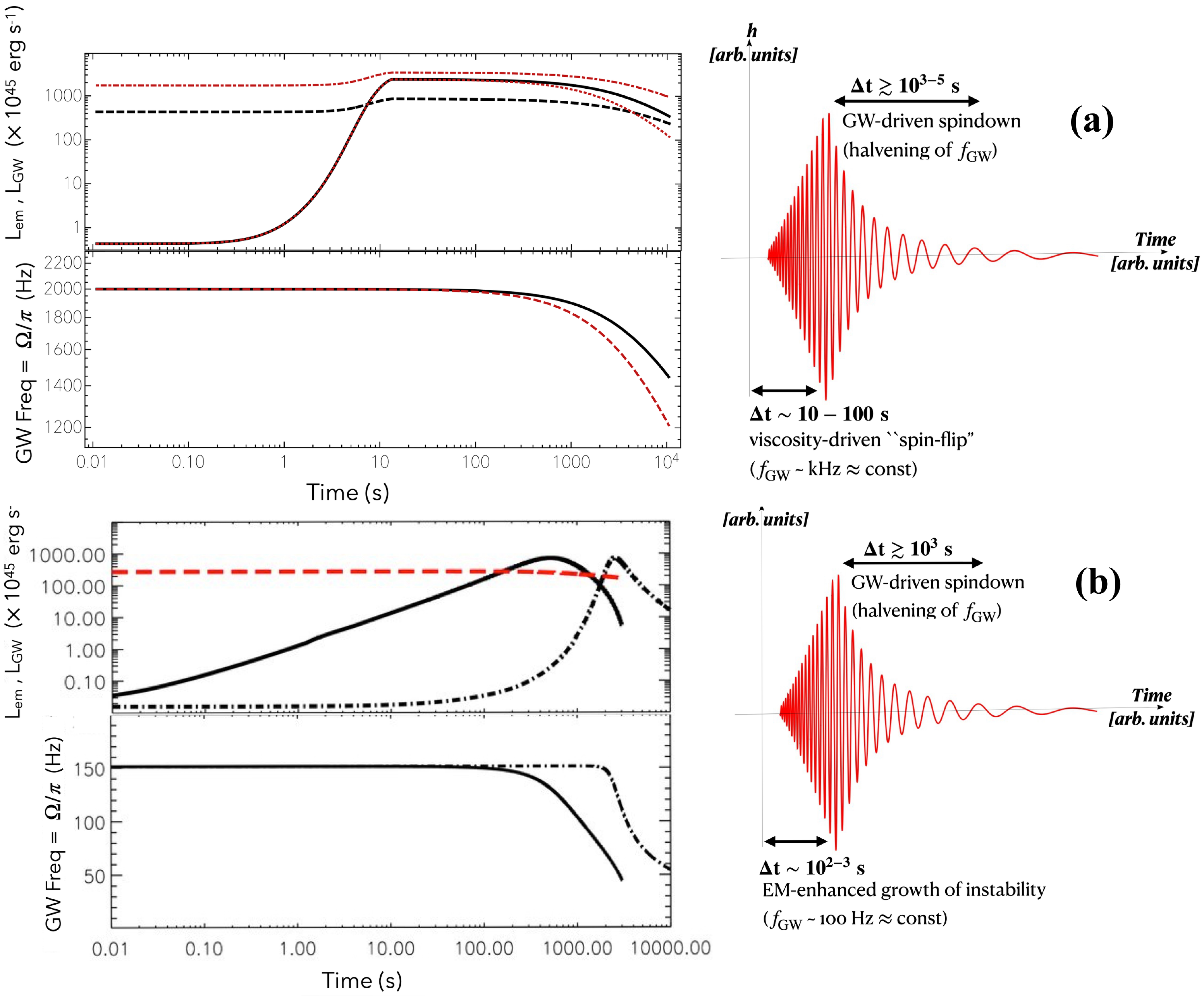}
\caption{(a) Spin-Flip. {\it Upper left:} the GW (continuous) and EM (dashed) spindown luminosity of a 1 ms-magnetar with initial tilt angle $\chi =3^\circ$, $\epsilon_{B, -3} =1$ and B$_{{\rm d}, 14}=1$ (black). GWs (dotted) and EM (dot-dashed) luminosity with a three times stronger dipole B-field and all other parameters fixed (red). The GW fast rise follows the growth of the tilt angle (see footnote 5). The EM emission grows as well, by a factor 2, due to a term $\left(1+ \sin^2 \chi \right)$ in Eq. \ref{eq:emspindown} (see footnote 4).  
{\it Lower left:} Signal frequency vs. time with the weaker (continuous, black) and stronger (dashed, red) dipole B-field. {\it Right panel:} Sketch of the time evolution of the GW signal (strain).~(b)~Bar-mode. {\it Upper left:} GW (continuous, black) and EM (dashed, red) luminosity of a NS with B$_{{\rm d}, 14}=1$ (adapted from \cite{Corsi09}). {\it Lower left:} signal frequency (twice the rotation frequency of the elliptical pattern) vs. time (from \cite{Corsi09}). {\it Right panel:} same as in (a).}
\label{fig:fig3}
\end{figure}

Therefore, a $\sim 10^3$~s long GRB X-ray plateau followed by a $\sim 10^3$ s GW signal~at $f <1$ kHz may represent the signature of a newborn magnetar with spin energy $E_{\rm rot} > 5 \times 10^{52}$ erg. The X-ray plateau would carry information about the NS spin~and dipole magnetic field (B$_{\rm d}$), while the GW signal tracks the $f$-mode amplitude ($\epsilon$) and its pattern frequency ($\Omega_{\rm p}$), in principle different from that associated to the plateau (\cite{Corsi09}).
The pattern rotation has a lower frequency ($\Omega_{\rm p}$) than $\Omega = \Omega_{\rm p} +\Lambda$:~this offers an advantage for GW signal detection, by placing its frequency in the range $\sim (0.1-1)$ kHz, where the Advanced LIGO/Virgo sensitivity is higher. For optimal NS parameters, the GW signal may reach a peak $h \sim 10^{-26}$ Hz$^{-1/2}$ at an horizon distance $d\sim 100$ Mpc for Advanced LIGO/Virgo with ideal matched-filter searches\cite{Corsi09}. 

In contrast to ideal matched-filter, realistic signal searches are strongly limited by computational resources and must be carried out with sub-optimal methods. 
Currently, four different dedicated pipelines exist for LIGO/Virgo data implementing four different detection techniques \cite{LIGO18}: STAMP (Stochastic Transient Analysis Multidetector Pipeline), HMM (Hidden Markov Model) tracking, ATrHough (Adaptive Transient Hough) and FreqHough (Generalized FrequencyHough). The discovery of the BNS merger GW 170817, at the distance of only 40 Mpc, provided a first performance test for such pipelines.  No NS signal was revealed, but it was shown with real data that the O2  horizon for the 4 pipelines ranged between 0.2 and 0.9 Mpc (\cite{LIGO18}), independent of the particular emission model. The foreseen sensitivity improvements in the years to come imply a realistic horizon $\sim 3$ Mpc for Advanced interferometers, and $\sim 20-30$ Mpc with 3rd generation detectors. 

To summarize this section, the huge reservoir of rotational energy of millisecond magnetars 
($\sim 10^{-2} M_\odot c^2$) 
and their extreme interior B-fields allow for strong GW signals to be emitted, potentially carrying orders of magnitude more energy than the GW signals expected from core-collapse or {\it ordinary} NS ($< 10^{-5} M_\odot c^2$; {\it e.g.} \cite{Saty09, Ander13}). 
This makes newborn magnetars potentially detectable in GWs over a much larger volume than any other isolated NS source ($> 10^{3.5}$ times larger), hence at a substantially higher rate despite their relative rarity. The shape of their GW signals carries the signature of the mechanism that has produced them, and provides insights into the physical properties of newborn magnetars.
\section{Fast Radio Bursts and millisecond magnetars} 
\label{sec:FRB}
Fast Radio Bursts (FRBs) are another class of astrophysical transients which has drawn much interest into young magnetars and the possibility that they are endowed with ultra-strong interior B-fields and large spin energy. 
FRBs are single, (sub)-millisecond radio pulses with GHz peak flux densities $S_\nu \sim$(0.1-10)~Jy and very large dispersion measures (the electron column density along the line of sight to the source) DM $\sim 100-2600$ pc cm$^{-3}$, 
which exceed the values in our galaxy. 
Assuming that the extra-DM is due to the intergalactic medium, typical comoving distances in the Gpc range, 
isotropic emitted energies of 
$E_{\rm iso} \sim 10^{38}-10^{40}$~erg, 
and brightness temperatures, $T_{\rm b} >10^{33}-10^{36}$~K were derived (see, {\it e.g.}, \cite{RaLor17, Katz18, Pet19, CoCha19, Zha20} for recent reviews). First discovered in Parkes archival data \cite{Lor07}, for long their nature remained uncertain, their astronomical origin 
being supported by (i) the precise scaling of the arrival times of individual pulses with frequency ($\propto \nu^{-2}$), as expected from radio waves propagating through a cold plasma  and (ii) the duration of the decaying tail of individual pulses vs. frequency ($\propto \nu^{-4.4}$), as expected from waves scattered by a medium with a Kolmogorov spectrum of inhomogeneities. Eventually, the discovery of the repeating FRB121102 \cite{Spit16}, \cite{Sch16} and the identification of its host galaxy at $z \approx 0.193$ \cite{Chat17} established FRBs as genuine extragalactic phenomena.  
More recently a second repeater was discovered (FRB180814; \cite{CHIME19a}), soon followed by seventeen new ones (\cite{CHIME19b, Fon20a}), indicating that the fraction of repeating FRBs may be much higher than originally thought.

Their short durations imply FRBs are associated to compact sources, {\it i.e.} 
the size of the emitting region should be 
$\lesssim c \Delta t \Gamma_2^2 \approx 3 \times 10^7~{\rm cm}~\Delta t_{-3} \Gamma^2$. In contrast~to GRB jets, where the pair-production opacity for $\gamma$-ray photons places a strong constraint on the bulk Lorentz factor, different models for FRBs have widely different requirements on the source bulk motion, from non-relativistic (e.g. \cite{Lyut20a, Lyut20b}) to extremely relativistic ($\Gamma \sim 10^2-10^3$ or even up to $10^5$ in shock models \cite{MaMeSi20, Belobod20}; see below).

Accounting for beaming  
$f_{\rm b} = \Delta \Omega/4 \pi$
within a solid angle $\Delta \Omega$, the energy 
released by FRBs is 
\be
\label{eq:energetic}
E_{\rm FRB} = E_{\rm iso} f_{\rm b} = \frac{S_\nu \Delta \nu \Delta t d^2_L}{1+z} f_{\rm b} \simeq 10^{38}~{\rm erg}~ \frac{S_\nu}{1~{\rm Jy}} \frac{\Delta \nu_9}{0.5} \Delta t_{-3}~d^2_{L, {\rm Gpc}} f_{\rm b}~\, ,
\ee
where $\Delta \nu_9$ is the bandwidth in GHz and the last equality is for a luminosity distance $d_L \approx 1$ Gpc ($z \approx 0.2$, or DM $\sim 300$ pc cm$^{-3}$; {\it e.g.} \cite{Pet19}). 

Besides FRB121102, the repeater  
FRB180916 is to date the only other source~of the class with an identified host galaxy and measured redshift,   
$z \approx 0.0337$ (150~Mpc; \cite{CHIME19b}), again consistent with its DM. This source displays also an enigmatic $\sim$ 16~d periodicity  in its bursting activity  \cite{CHIME20per, Zh20, Le20}. 
FRB121102 has a persistent radio source counterpart with luminosity $\nu L_\nu \approx 10^{39}$ erg s$^{-1}$ at 1.4 GHz and a flat spectrum $\propto \nu^{-0.2}$ up to $\sim 10$ GHz beyond which it steepens to $\alpha \sim -1$.  Based on its emission properties an estimated age $\gtrsim 30$ yrs was derived for the persistent radio source, hence for FRB 121102's progenitor \cite{Pi16, Metz17}. 

The puzzling features of FRBs boosted a flurry of theoretical models invoking compact object progenitors (see \cite{Pla19}). If all FRBs are produced by 
a single type of progenitors,
the existence of repeaters demonstrates that, despite their energy, FRBs are not produced in one-time-only catastrophic events ({\it e.g.} a NS collapsing to a BH).
NSs play a major role in FRB models, 
either  as isolated sources of magnetically-powered flares  or as members of binary systems (see, {\it e.g.}, \cite{Hess18, Gou20}).

The progenitors of FRBs must meet the constraint of their estimated all-sky rate $\sim 10^{3-4}$ events/day, or a rate density 
$\sim  10^{4-5} f^{-1}_{\rm b}$ Gpc$^{-3}$ yr$^{-1}$ up to $z\approx 0.5$ ($\sim 36$ Gpc$^3$). If repeating FRBs dominated the population, as suggested by the many already discovered, a much lower birth rate of FRB sources would be implied. A burst rate $\dot{N}_0 \sim$~1 per day is estimated for FRB121102 \cite{LuKu16}, while the estimated age of this source suggests an active lifetime $\tau_{\rm active} \gtrsim 30$ yr. If these values were typical of FRBs, each repeater would produce $N_{\rm FRB} \sim 10^4 \dot{N}_0 \left(\tau_{\rm active}/30{\rm yrs}\right)$ FRBs and the corresponding source birth rate would be 
${\cal R} \sim (1-10) f_{\rm b}^{-1} \dot{N}_0^{-1} \left(\tau_{\rm active}/30~{\rm yrs}\right)^{-1}$~ 
Gpc$^{-3}$~yr$^{-1}$. This is a small fraction of the rate of core-collapse supernovae (CCSN) $\sim 3 \times 10^5$ Gpc$^{-3}$ yr$^{-1}$(\cite{MaDi14}) and roughly comparable to the SLSN and (beaming-corrected) GRB rates \cite{LuKu16, LuKu18, NichollEA17}. Overall, a small fraction of all CCSN remnants broadly fit into this picture, possibly a small subset of the magnetar population \cite{Marg20}. 

A digression on ordinary radio pulsars is in order here. Their NSs are known to produce highly collimated beams of coherent radio waves with very large brightness temperatures, 
\be
\label{eq:Tb}
T_{\rm b} = \frac{S_\nu}{2 k_{\rm B}} \left(\frac{c d}{\nu R}\right)^2 \gtrsim \frac{S_\nu}{2 k_{\rm B}} \left(\frac{d}{\nu \Delta t}\right)^2 \approx 3 \times10^{20}\frac{S_\nu}{{\rm mJy}}\left(\frac{d_{\rm kpc}}{\nu_9 \Delta t_{-3}}\right)^2~{\rm K} \, ,
\ee 
$k_{\rm B}$ being Boltzmann's constant and $d$ the distance. While the radio mechanism~in pulsars is not fully understood yet, it is well established that it is powered by the~NS electromagnetic spin down (Eq. \ref{eq:emspindown}) with a characteristically low \cite{GJ69} radio efficiency\footnote{Normalized to the spindown power of the Crab pulsar and an acceleration potential $\varphi \sim 10^{12}$ V.},  $\epsilon_r \sim 10^{-5} \gamma_6 \left({L}_{{\rm EM}, 38}/4.6\right)^{-1/2}$, primarily limited by the ability of the~NS magnetosphere to accelerate a large number of electrons to relativistic energies~($\gamma \sim 10^6$).

Some radio pulsars sporadically emit Giant Pulses (GPs) with very short duration ($\sim \mu$s) and  much larger fluence than ordinary radio pulses. GPs can be interpreted as highly beamed, rotation-powered pulses \cite{CoWa16}. The brightest GP from the Crab pulsar had a 
peak fluence $\sim 2$ MJy at $\sim$ 9 GHz (2.5 GHz bandwidth; \cite{CoWa16}), implying that $\sim 10^{28}$ erg were released in its $\sim 0.4$ ns duration ($T_{\rm b} \sim 10^{41}$ K). By virtue of Eqs. \ref{eq:energetic} and \ref{eq:Tb} this requires $\epsilon_r/f_{\rm b} < 0.05$, or $f_{\rm b} > 2 \times 10^{-4}$ for $\epsilon_r \sim 10^{-5}$, in order not to exceed the spindown power  
being transferred to relativistic particles. 
Even smaller $f_{\rm b}$ values may in general be expected of coherent radiation, in principle allowing for brighter rotationally-powered GPs from the Crab pulsar \cite{CoWa16}. 

These ideas become challenging when extended to interpret FRBs as extreme~versions of spindown-powered GPs. Indeed, FRBs typically release $> 10^8$ times~more energy than the brightest GP, which can't be 
ascribed to beaming alone. 
 Moreover,~a rotation-powered burst with beaming $f_{\rm b}$ and duration $\Delta t$ would have 
fluence 
\be
\label{eq:fluence}
F_\nu = S_\nu \Delta t \approx 
 8 \left(\epsilon_r / f_{\rm b} \right) L_{{\rm EM},40}~ \Delta t_{-3} \left(\Delta \nu_9\right)^{-1} d^{-2}_{\rm Gpc}~~{\rm mJy~ms}   \, 
 \ee
which, for 
a typical detection threshold of  $\sim$ 1 Jy ms, 
requires $\left(\epsilon_r/f_{\rm b}\right) L_{{\rm EM}, 40}~ \Delta t_{-3} \gtrsim 10^2 \Delta \nu_9 ~d^2_{\rm Gpc}$. Recall that 
$\epsilon_r / f_{\rm b} \sim 10^{-2}$ in the Crab's brightest GP and $\epsilon_r / f_{\rm b} \lesssim 10^{-3}$ for its more typical GPs; 
even  allowing for a large $\epsilon_r / f_{\rm b} =1$ in FRBs and a small bandwidth $\Delta \nu_9 \approx 0.1$, L$_{{\rm EM}, 40} \gtrsim 10$  would still be 
required, which only holds~for a young NS of age (Eq. \ref{eq:emspindown}) 
\be
\label{eq:time}
t  \lesssim I_{45}{R^{-3}_6} \left[5.5 \left(B_{{\rm d,} 14}~ d_{\rm Gpc}\right)^{-1} \left(\epsilon_r/f_b\right)^{1/2} - 0.0045 P^2_{-3}B^{-2}_{{\rm d,} 14} R^{-3}_6 \right]~{\rm yrs}\, ,
\ee
where the spin period and B-field refer to birth values. The limit in Eq. \ref{eq:time} is~consistent with the estimated age of FRB~121102, {\it i.e.} $\gtrsim 30$ yr, only for P$_{-3} < 8$ and B$_{{\rm d,} 14} < 0.2$.  
Rotation-powered FRB models may thus be viable only for~young, ms-spinning NSs with moderately strong dipole B-fields (cf. \cite{Lyu16, MeETAL18}) under favourable assumptions on the other parameters in Eq. \ref{eq:time}.

Due to these difficulties, the magnetic field was considered as an alternative~energy source for NS-powered FRBs. Moreover, their characteristic range of energies and timescales, as well as the bursting activity of repeating FRBs, are reminiscent, broadly speaking, of the~X/$\gamma$-ray behaviour of classical magnetars ({\it e.g.}, \cite{KaBel17, EspositoEA18}). This led several authors~to propose extragalactic magnetars as possible/likely progenitors of FRBs ({\it e.g.} \cite{Lyub14, Kul15, Kaz16, Bel17, Metz17, LuKu18, LuKu19, Metz19, Bel20, Lyut20a, Lyut20b}). In this scenario, FRBs are powered by the release of the magnetic energy reservoir of their progenitors, which is $\gg 10^{47}$~erg and thus largely sufficient to power FRB activity for very long times, in analogy with the X/$\gamma$-ray bursts/flares of galactic magnetars. 
Several mechanisms are proposed for coherent radio emission of $\sim 10^{36}-10^{40}$ erg on a $\sim$ ms timescale:~(a)~ magnetospheric models envision particle bunches emitting curvature radiation in a very~strong~B-field close to the NS (\cite{LuKu18, LuKu19, Lyu20a, Lyut20a, Lyut20b}); (b) external models point to~a synchrotron maser, excited in the interaction of a flare-induced plasma outflow~with a surrounding highly magnetised wind, produced by previous flares or by the magnetar spindown, {\it i.e.} a magnetar wind nebula \cite{Lyub14, Kul15, Bel17,  Metz17, Metz19}.

Observationally, rotation-powered and magnetically-powered models have proven difficult to disentangle. So far polarization and rotation measure studies ({\it e.g.} \cite{Mas15, Petr17, Caleb18, Mic18}), aimed at revealing signatures of highly magnetized progenitors,~have not led to firm conclusions.  On the other hand, the
lack of energetic radio flares from AXPs/SGRs represented for long a  
criticality for magnetar-based scenarios. 
Two very intense radio pulses were reported in association with X-ray bursts from a transient galactic magnetar\footnote{Israel et al. (2020), submitted to ApJ.}, but their energy was down by $\sim$ 5 orders of magnitude even~with respect to the 
weakest known FRB (FRB 180916 which released $E_{\rm iso} \sim 5 \times 10^{36}$~erg; \cite{CHIME19b}). The situation changed dramatically with the recent discovery of a bright radio flare temporally coincident with an X-ray burst in the classical magnetar SGR 1935+2154 \cite{Boch20, CHIME20, Merd20}. The event contained two 0.5-ms sub-bursts separated by a quiescent period of $\sim$ 28 ms. Its fluence, constrained by STARE observations to be $> 1.5$ MJy ms at 3 GHz, implies an isotropic-equivalent energy $\sim 10^{35}$ erg at a distance $\sim 9.5$ kpc \cite{Boch20}, just a factor $\sim 40$ less than FRB 180916. This crucial discovery demonstrated that: (a) magnetars are capable of producing~radio~bursts~with similar properties to those of extragalactic FRBs; 
(b) magnetar FRBs can be  
associated~to~X-ray bursts, as envisaged in many of the proposed models; (c) the radio bursts are almost exactly simultaneous with the peaks of the X-ray bursts, favoring models~in which the emission in the two bands is co-located, either both produced in~the~NS magnetosphere or in the shock region within the magnetar wind nebula; (d) the FRB-like flare of SGR 1935+2154 could not be powered by the instantaneous spindown luminosity ($\sim 10^{35}$ erg s$^{-1}$) of the NS. A mechanism for the long-term accumulation of rotation power, capable of releasing it on a $\sim$ ms timescale, e.g. via some kind of instability\footnote{NS glitches represent an example of this kind of processes.}, may still be consistent with the energetics of the event; (e) the radio-to-X-ray fluence ratio $\sim 10^{-5}$ \cite{CHIME20, Marg20}   
was, at least in this event, within the 
relatively narrow range predicted by 
external shock models (e.g. \cite{MaMeSi20}), and in contrast to the expectations from other models (see \cite{Ravi20} for a thorough discussion).

The repeater FRB 121102, which has been emitting sporadic bursts since its discovery \cite{Opp17}, offers a unique test of the energetic requirements in the magnetar scenario.   In \cite{Spit14} 11 bursts were studied, which released a total isotropic-equivalent $\sim 4.5 \times 10^{39}$ erg. With a total on-source time $\sim$ 15.8 hrs, these imply a long-term average luminosity\footnote{The same result is obtained using VLA or Green Bank data (\cite{LuKu18}).} \cite{LuKu18}
\be
\langle L_{\rm FRB} \rangle \approx 8 \times 10^{34}~ \left(f_{\rm b, tot}/\epsilon_r\right)~{\rm erg~s}^{-1} \, ,
\ee
where $f_{\rm b, tot} = 11 \times \langle f_{\rm b} \rangle$ 
is the sum of the beaming factors of the 11 bursts. Multiplying $\langle L_{\rm FRB} \rangle$ by $\tau_{\rm active} \gtrsim 30$ yrs gives a minimal energy $E_{\rm min} \sim 8 \langle f_{\rm b} \rangle / \epsilon_r \times 10^{44}$ ergs, 
which translates to a minimal B-field 
$B_{\rm min} = \left(6\, E_{{\rm min}}/R_*^3\right)^{1/2} \approx 6 \times 10^{13} \left(f_{\rm b} / \epsilon_r \right)^{1/2}$~G.~Thus, if FRB 121102 hosts a magnetar, values of $f_{\rm b}/\epsilon_r \gtrsim 10^4-10^5$ would imply a magnetic energy $\sim 10^{49}-10^{50} \left(t_{\rm active}/30{\rm~ yr}\right)$ erg, or a minimum $B_{\rm int}\sim(0.6-2) \times 10^{16} \left(t_{\rm active}/30{\rm yr}\right)^{1/2}$ G. 
As discussed above, $f_{\rm b}/\epsilon_r \sim 10^{2}-10^{3}$ is estimated in radio pulsar GPs. In the magnetar SGR 1935+2154, on the other hand, the measured radio-to-X-ray fluence suggests $f_{\rm b}/\epsilon_r \sim 10^{4}-10^{5}$ if the X-rays were not beamed, or only mildly beamed.

An independent test of the magnetar scenario is based only on the persistent~radio counterpart of FRB 121102. With a luminosity $\nu L_\nu \sim 10^{39}$ erg s$^{-1}$ and flat spectrum up to $\sim 10$ GHz, beyond which a turnover appears, the source is straightforwardly interpreted as a synchrotron nebula, with size $R <  0.7$ pc and $\tau_{\rm age} \sim 30-100$ yrs  \cite{Metz17}. Its radio luminosity is consistent with the spindown power of a young, moderately magnetized and ms-spinning NS (Eq. \ref{eq:time}).  However, its spectral properties - in particular the peak at $\sim 10$ GHz - can only be explained with a number of emitting particles  
${\cal N} \gtrsim 10^{52} \left(t_{\rm age}/30~{\rm yr}\right)^{2/3}$, 
too large to be provided by the typical mechanism 
of pulsar wind nebulae\footnote{The spectral peak at $\nu \sim 10$ GHz and the luminosity $L_\nu \sim 10^{29}$ erg s$^{-1}$ Hz$^{-1}$ below the peak lead to the conclusion that the energy of the emitting particles peaks at $\gamma_e \sim 10^2/B^{1/2}$, where $B$ (in gauss) is the magnetic field strength in the emitting region, and that ${\cal N} B \sim 2 \times 10^{50}$ G (\cite{Bel17}). Moreover, the constraint that the particle cooling break $\nu_c = 10 e m_e c/\left(\sigma^2_T B^3 t^2\right)$ be $>10$ GHz implies $B < 0.03 \left(t_{\rm age}/30~{\rm yr}\right)^{-2/3} (\nu_{c,9}/10)^{-1/3}$ G, from which ${\cal N} > 7 \times 10^{51} t^{2/3}_9 (\nu_{c, 9}/10)^{1/3}$ is derived. Particles are accelerated at the NS surface to relativistic energies, giving rise to an electrical current $i = \mu \Omega^2/c$ and leading to copious pair-production in the magnetosphere. Pairs are produced at the rate $\dot{N}^{\pm} \sim 2 {\cal M} i/e$, where the multiplicity ${\cal M}$ is the number of $e^\pm$ pairs per accelerated particle, and typically ${\cal M} \lesssim 10^3$ in PWNe. Because L$_{\rm EM} \sim i^2/c$, the NS in FRB121102 should produce pairs with a multiplicity ${\cal M} \sim 10^7-10^{10}$ in order to inject $10^{52}$ particles, given the spindown luminosity $\sim 10^{39}$ erg s$^{-1}$. This is several orders of magnitude larger than typical in PWNe, pointing to a different mechanism or to very different conditions existing in this source.}  (PWN) with the current spindown power \cite{Bel17}. 
${\cal N} \gtrsim 10^{52}$ may instead  be provided in two possible ways: \\
(i) in the early stages of the nebula lifetime, when $R$ was much smaller and its large compactness $\ell = L_{\rm EM} \sigma_T/(R m_e c^3)$ allowed a much more efficient pair-production than typical of PWNs (\cite{Metz17, Bel17}). Following \cite{Bel17} two possible solutions can be found which would provide $\sim 10^{52}$ particles to the nebula well before its current age. One is a moderately magnetized  NS, with B$_{\rm d} \gtrsim 10^{13}$ G and birth spin P $\gtrsim$ 6 ms, the other a B$_{\rm d} \gtrsim 10^{15}$ G NS with any P$\sim (1- 40)$ ms. \\
(ii) during the  whole $\gtrsim 30$ yrs lifetime of the remnant, if the central engine has continuously released magnetic energy through flares at its current rate (assuming  $f_{\rm b}/\epsilon_r > 10^4$), and if each flare 
released radiation (in erg) and particles (electrons) with a similar ratio $N_{\rm part}/{E}_{\rm rad} \gtrsim 10^2$ erg$^{-1}$ to that estimated in the Giant Flare from the classical magnetar SGR 1806-20 \cite{Bel17}.

In either case the constraints from the persistent radio source agree with those obtained from the FRBs indicating  that a highly magnetized 
NS  is a likely central engine in FRB 121102, and further suggesting it was spinning at $\sim$ millisecond period at birth.
\section{Conclusions}
Our knowledge of GRBs and SNe, including SLSNe, has advanced a great deal in the last two decades; the pace of discovery in GW and FRB studies has been very fast in recent years. In all these areas, models involving millisecond spinning magnetars appear to be promising for interpreting paroxysmal phenomena that characterise these sources. Direct evidence for the presence of such extremely powerful central engines is still lacking. 

An unambiguous signature of newborn magnetars would be the detection of fast, rapidly-evolving periodic signals arising from the star rotation. Searches in the electromagnetic channels face serious difficulties. It is hard to image ways in which a newborn magnetar's signal could emerge out of the huge matter depth of an expanding SN envelope, or out of the relativistic shocks (or even an $\sim$ hr-old magnetar wind nebula) of a GRB. Recurrent FRBs from repeaters, should they occur at fixed rotational phases, may allow detection of the underlying magnetar periodic signal in a manner similar to that of Rotating Radio Transients (RRaTs) \cite{ZhangEA18}.
As GW signals are intrinsically locked to the star rotation and travel unimpeded they provide better prospects for revealing the characteristic signal of newborn, fast spinning magnetars. 
The detectability of  their hrs-long, time-reversed chirps
is presently limited to an horizon of $\sim 3$~Mpc by a combination of detector sensitivity 
and computing power.

Research in these frontier areas will likely be thriving 
even faster in the years to come, both through 
observations and theoretical studies;  
a wealth of new results and, possibly, some surprises
are to be expected. 

\section{Acknowledgements}
We wish to thank G. Ghisellini, B. Margalit, B. Metzger and A. Papitto for careful reading of the manuscript and helpful comments and suggestions. LS acknowledges financial contributions from ASI-INAF agreements 2017-14-H.O and  I/037/12/0 and from ``iPeska" research grant (P.I. Andrea Possenti) funded under the INAF call PRIN-SKA/CTA (resolution 70/2016).
 
\bibliographystyle{spphys.bst}
\bibliography{Magnetar_biblio.bib}
\end{document}